\documentclass[a4paper, 11pt]{article}

\usepackage{a4wide}
\usepackage{latexsym}
\usepackage{graphicx}
\usepackage{color}
\usepackage[hypertex]{hyperref}

\title{A Discrete Evolutionary Model for Chess Players' Ratings}

\author{Trevor Fenner, Mark Levene, and George Loizou \\
Department of Computer Science and Information Systems \\
Birkbeck, University of London \\
London WC1E 7HX, U.K. \\ \{trevor,mark,george\}@dcs.bbk.ac.uk}

\date{}

\begin{document}

\maketitle

\newtheorem{theorem}{Theorem}[section]
\newtheorem{corollary}[theorem]{Corollary}
\newtheorem{lemma}[theorem]{Lemma}
\newtheorem{proposition}[theorem]{Proposition}
\newtheorem{definition}{Definition}[section]
\newtheorem{algorithm}{Algorithm}
\newtheorem{example}{Example}[section]

\begin{abstract}

The Elo system for rating chess players, also used in other games and sports, was adopted by the World Chess Federation over four decades ago. Although not without controversy, it is accepted as generally reliable and provides a method for assessing players' strengths and ranking them in official tournaments.

It is generally accepted that the distribution of players' rating data is approximately normal but, to date, no stochastic model of how the distribution might have arisen has been proposed. We propose such an evolutionary stochastic model, which models the arrival of players into the rating pool, the games they play against each other, and how the results of these games affect their ratings. Using a continuous approximation to the discrete model, we derive the distribution for players' ratings at time $t$ as a normal distribution, where the variance increases in time as a  logarithmic function of $t$. We validate the model using published rating data from 2007 to 2010, showing that the parameters obtained from the data can be recovered through simulations of the stochastic model.

The distribution of players' ratings is only approximately normal and has been shown to have a small negative skew. We show how to modify our evolutionary stochastic model to take this skewness into account, and we validate the modified model using the published official rating data.

\end{abstract}


\section{Introduction}

The Elo system for rating chess players \cite{ELO86}, named after its creator Arpad Elo, has been employed by the World Chess Federation for over four decades as a method for assessing players' strengths and ranking them in official tournaments. Although not without controversy, it is accepted as generally reliable, and is also used in other games and sports such as Scrabble, Go, American football and major league basketball.

\smallskip

The Elo rating system is based on the model of paired comparisons \cite{DAVI88}, which can be applied
to the problem of ranking any set of objects for which we have a preference relation.
The model is particularly useful in that a ranking can be obtained in situations where a preference exists only for some of the pairs of objects under consideration. Paired comparison models have been successfully applied to measure ability in competitive games and sports \cite{JOE91,GLIC99a}, the most notable example being the widely used Elo system for rating chess players.

Several extensions to the Elo system have been proposed, notably the Glicko \cite{GLIC99a} and TrueSkill \cite{HERB06} Bayesian rating systems. Both these systems estimate, in addition to the rating, the degree of uncertainty that the rating represents the player's true ability. The uncertainty allows the system to control the change made to the rating after a game has been played. In particular, if the uncertainty is low then the changes made to the rating should be smaller as the rating is already reasonably accurate, while if the uncertainty is high then the changes made to the rating should be larger.

\smallskip

Here we adopt the Bradley-Terry model \cite{BRAD52}, which provides the theoretical underpinning of Elo's model, where the probability $p_{\alpha \beta}$ that a player $A$, whose strength is $\alpha$, wins against a player $B$, whose strength is $\beta$, is given by the logistic function $L_C(\cdot)$, namely
\begin{equation}\label{eq:log}
p_{\alpha \beta} =  L_C (\alpha - \beta) = \frac{1}{1 + \exp \left(- C(\alpha - \beta) \right) },
\end{equation}
where $C$ is a positive scaling factor.
We note that $L_C(\cdot)$ is strictly monotonically increasing,
$\lim_{x \to -\infty} L_C(x) = 0$, $\lim_{x \to +\infty} L_C(x) = 1$ and $L_C(0) = 0.5$.
Moreover,
\begin{equation}\label{eq:sum}
L_C(x) + L_C(-x) = 1.
\end{equation}


\medskip

In this paper we are interested in the distribution of ratings within the pool of
players that arises as a result of the model induced by (\ref{eq:log}). We are not aware of any research
in this direction, although it is generally accepted that this distribution is well approximated by a Gaussian (i.e. normal) distribution \cite{CHAR96,BILA09}. It is worth mentioning that Elo \cite{ELO86} claimed that the distribution of ratings of established chess players was not Gaussian, and suggested the Maxwell-Boltzmann distribution as an alternative that fitted the data he used slightly better.

\medskip

The rest of the paper is organised as follows.
In Section~\ref{sec:elo} we review the Elo rating system, and in Section~\ref{sec:fide} we do some exploratory data analysis on published official chess rating data. We show that the Gaussian distribution provides a very good fit to the data, but there is a small negative skew present.
In Section~\ref{sec:urn} we propose an evolutionary stochastic model, which as a first attempt assumes a symmetric distribution of ratings. The derivation of the distribution is presented in Section~\ref{sec:derive}, where we prove that the resulting distribution is indeed normal, with the interesting feature that the variance increases with time in a logarithmic fashion.
In Section~\ref{sec:chess} we validate the model using published rating data from January 2007 to January 2010, and
in Section~\ref{sec:skew} we modify the model to allow for the skewness present in the data. With reference to this data, we show through simulation that the modified model yields a better approximation to the actual distribution. Finally, in Section~\ref{sec:conc} we give our concluding remarks.

\section{Elo's Rating System}
\label{sec:elo}

We now summarise Elo's rating system \cite{ELO86} in order to set in context the evolutionary model that we present in Section~\ref{sec:urn}.

\smallskip

The fundamental assumption of Elo's rating system is that each player has a current playing strength.
In a game played between players $A$ and $B$, with unknown strengths $\Phi_A$ and $\Phi_B$, the score of the game for player $A$ is denoted by $S_{AB}$, where $S_{AB}$ is 1 if $A$ wins, 0 if $A$ loses and $0.5$ if the game is a draw.
Its expected value is assumed to be \cite{GLIC99b}
\begin{equation}\label{eq:elo}
E(S_{AB}) = L_C(\Phi_A - \Phi_B),
\end{equation}\label{eq:elo-factor}
where $E(\cdot)$ is the expectation operator and
\begin{equation}
C = \frac{\ln 10}{400} \approx 0.0058.
\end{equation}
\smallskip

The Elo system attempts to estimate the strength $\Phi_A$ of player $A$ using a calculated rating $R_A$, which is adjusted according to the results of games played by $A$. We observe that this model is related to the Bradley-Terry model for paired comparison data \cite{BRAD52}; see also \cite{DAVI88}.

After playing a game against player $B$, player $A$'s rating is adjusted according to the following formula (see equation (2) in \cite{GLIC99b})
\begin{equation}\label{eq:adjust}
new \ R_A = old \ R_A + K (S_{AB} - E(S_{AB})),
\end{equation}
where $K$ (known as the $K$-factor) is the maximum number of points by which a rating can be changed as a result of a single game. (A high $K$-factor gives more weight to recent results, while a low $K$-factor increases the relative influence of results from earlier games.) In the Elo system the $K$-factor is typically between 10 and 30.
(There has been some controversy involving a recent proposal by the World Chess Federation to change the $K$-factor \cite{SONA09,ZULT09}.) For the purpose of experimentation we have fixed the $K$-factor at 20.

\smallskip

When using (\ref{eq:adjust}) to update $R_A$, $E(S_{AB})$ is estimated from (\ref{eq:elo}) using the current values of $R_A$ and $R_B$ as estimates of $\Phi_A$ and $\Phi_B$, respectively.

Player $B$'s rating is updated similarly. We note that, after updating both $A$'s and $B$'s ratings, the sum of their ratings remains unchanged. The above method can be straightforwardly extended to the case of a player competing in a tournament, or to a number of games played over a given period.

\section{The Distribution of Elo Rating Data}
\label{sec:fide}

The World Chess Federation, known as FIDE, publishes a rating list several times each year. Traditionally FIDE published the rating list every three months, but from 2009 has moved to bi-monthly publication; the official rating data can be obtained from \url{http://ratings.fide.com}.

\smallskip

Here we are interested in the distribution of the players' ratings. It has been confirmed by Charness and Gerchak \cite{CHAR96}, and by Bilali\'c et al. \cite{BILA09} that the distribution is well approximated by a Gaussian distribution. We recall that the probability density function for a Gaussian random variable $X$ takes the form,
\begin{equation}\label{eq:gauss}
\frac{1}{\sqrt{2 \pi \sigma^2}} \exp \left( \frac{-(X - \mu)^2}{2 \sigma^2} \right),
\end{equation}
where $\mu$ is the mean and $\sigma$ is the standard deviation of $X$.

\smallskip

\begin{figure}
\begin{minipage}{0.49\textwidth}
\centering{\includegraphics[scale=0.40]{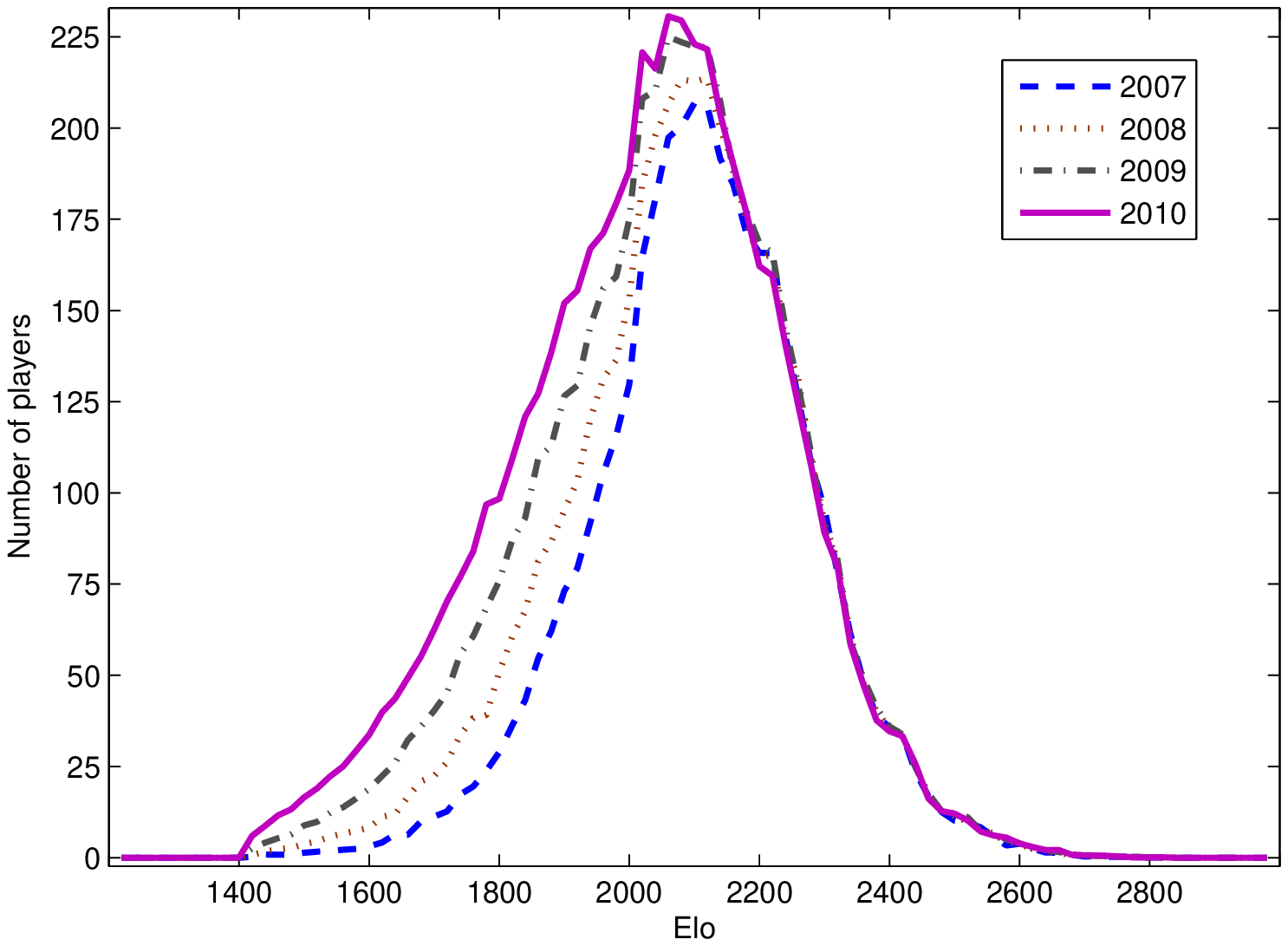} } \\
\end{minipage}
\begin{minipage}{0.49\textwidth}
\centering{\includegraphics[scale=0.40]{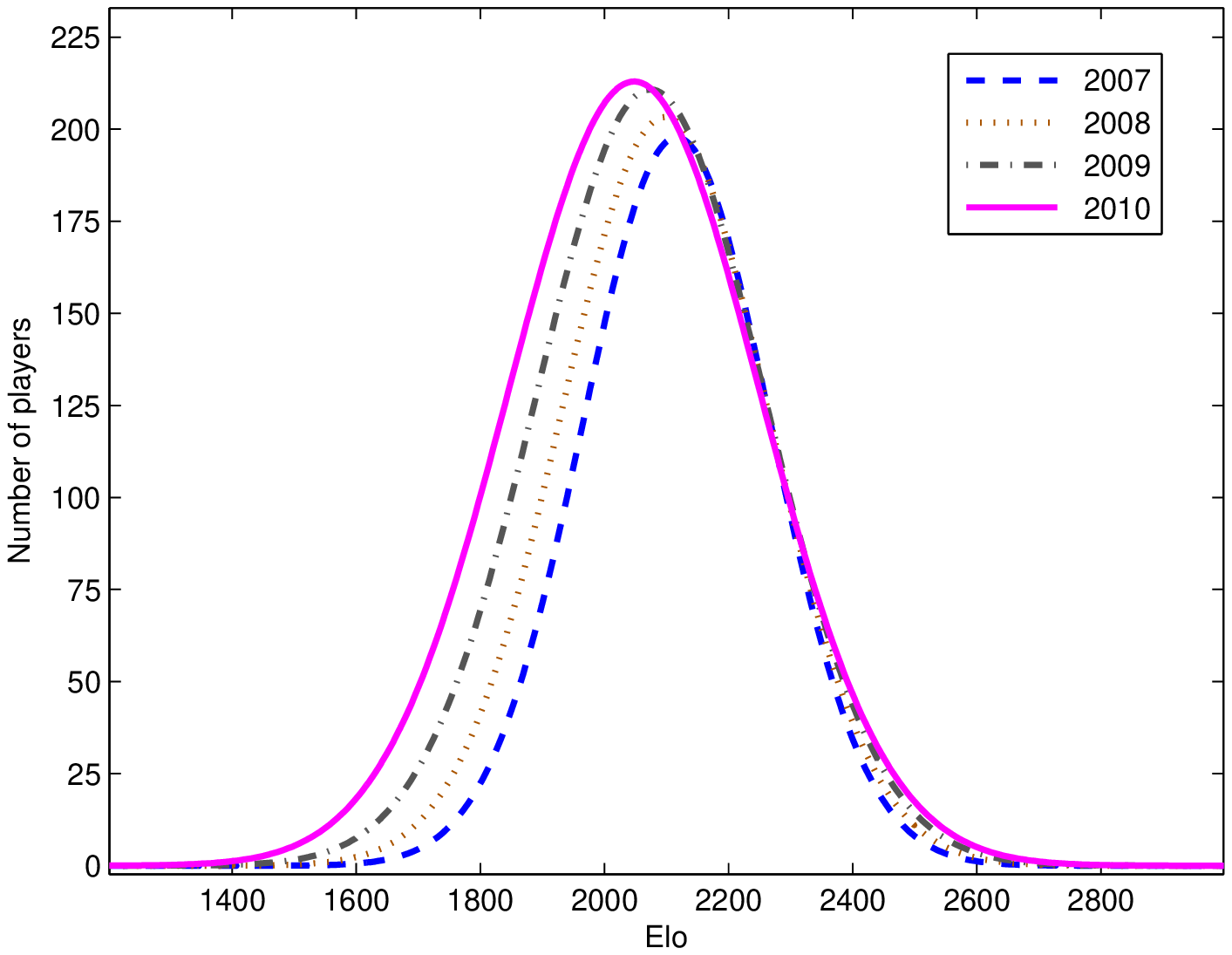} } \\
\end{minipage}
\vspace*{8pt}\caption{\label{fig:rating} Plots of the binned rating data (left) and the fitted Gaussians (right) from January 2007 to January 2010}
\end{figure}

With these observations in mind, we performed some exploratory data analysis on the FIDE rating data from January 2007 to January 2010. To test the normality of the data, we binned each of the four data sets, taking the bin width to be 20 (the fixed $K$-factor). The resulting plots for January 2007 to January 2010 are shown on the left-hand side of Figure~\ref{fig:rating}.

\smallskip

We then fitted a constant multiple of a Gaussian distribution to each of the four data sets, using Matlab.
The plots for the fitted data for January 2007 to January 2010 are shown on the right-hand side of Figure~\ref{fig:rating}.
The fitted parameters, $Q$, $\mu$ and $\sigma$, are shown in Table~\ref{table:gauss}, where $Q$ is the multiplicative constant.
Clearly, $Q$ is an approximation to the actual number of players $P$.
Table~\ref{table:gauss} also shows $R^2$, the coefficient of determination \cite{MOTU95}. It can be seen that this is close to 1, which indicates a very good fit. For comparison, the last two columns in the table show the mean $\widehat{\mu}$ and standard deviation $\widehat{\sigma}$ computed from the actual FIDE rating data. On average $\widehat{\sigma}$ is about 7-15 Elo points greater than the fitted standard deviation $\sigma$.

\smallskip

\begin{table}[ht]
\begin{center}
\begin{tabular}{|c|c|c|c|c|c|c|c|}\hline
Year & $Q$    & $\mu$    & $\sigma$ & $R^2$  & $P$    & $\widehat{\mu}$ & $\widehat{\sigma}$ \\ \hline
2007 & 75167  & 2096.400 & 151.604  & 0.9936 & 77056  & 2100.127        & 166.203 \\
2008 & 84844  & 2077.400 & 166.452  & 0.9908 & 87075  & 2073.566        & 181.918 \\
2009 & 97070  & 2034.000 & 183.706  & 0.9859 & 99223  & 2044.687        & 196.639 \\
2010 & 107874 & 2007.600 & 202.092  & 0.9815 & 109373 & 2015.650        & 209.622 \\ \hline
\end{tabular}
\end{center}
\caption{\label{table:gauss} The parameters for the fitted Gaussians in Figure~\ref{fig:rating}}
\end{table}
\smallskip


It can be seen that the plots on the left-hand side of Figure~\ref{fig:rating} appear to show a
small negative skew. (We note that this is in contrast to the positive skew of the Maxwell-Boltzmann distribution, suggested by Elo \cite{ELO86}.) As a next step, we therefore investigated the skewness of the data for 13 rating periods from October 2006 to September 2009. The skewness $s$ is defined by
\begin{displaymath}
s = \frac{E(X - \mu)^3}{\sigma^3}.
\end{displaymath}

\begin{figure}
\begin{minipage}{0.49\textwidth}
\centering{\includegraphics[scale=0.40]{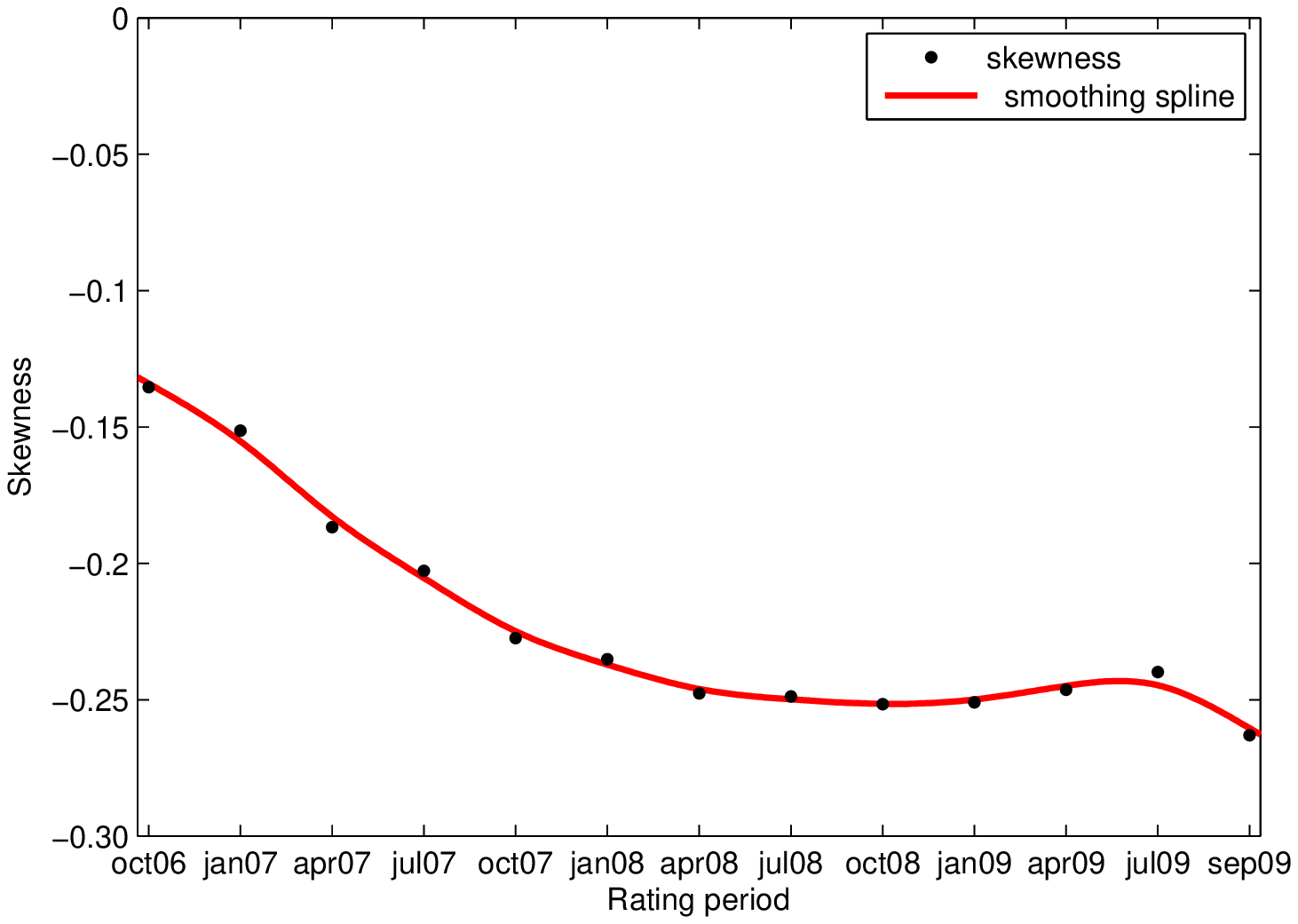} } \\
\end{minipage}
\begin{minipage}{0.49\textwidth}
\centering{\includegraphics[scale=0.40]{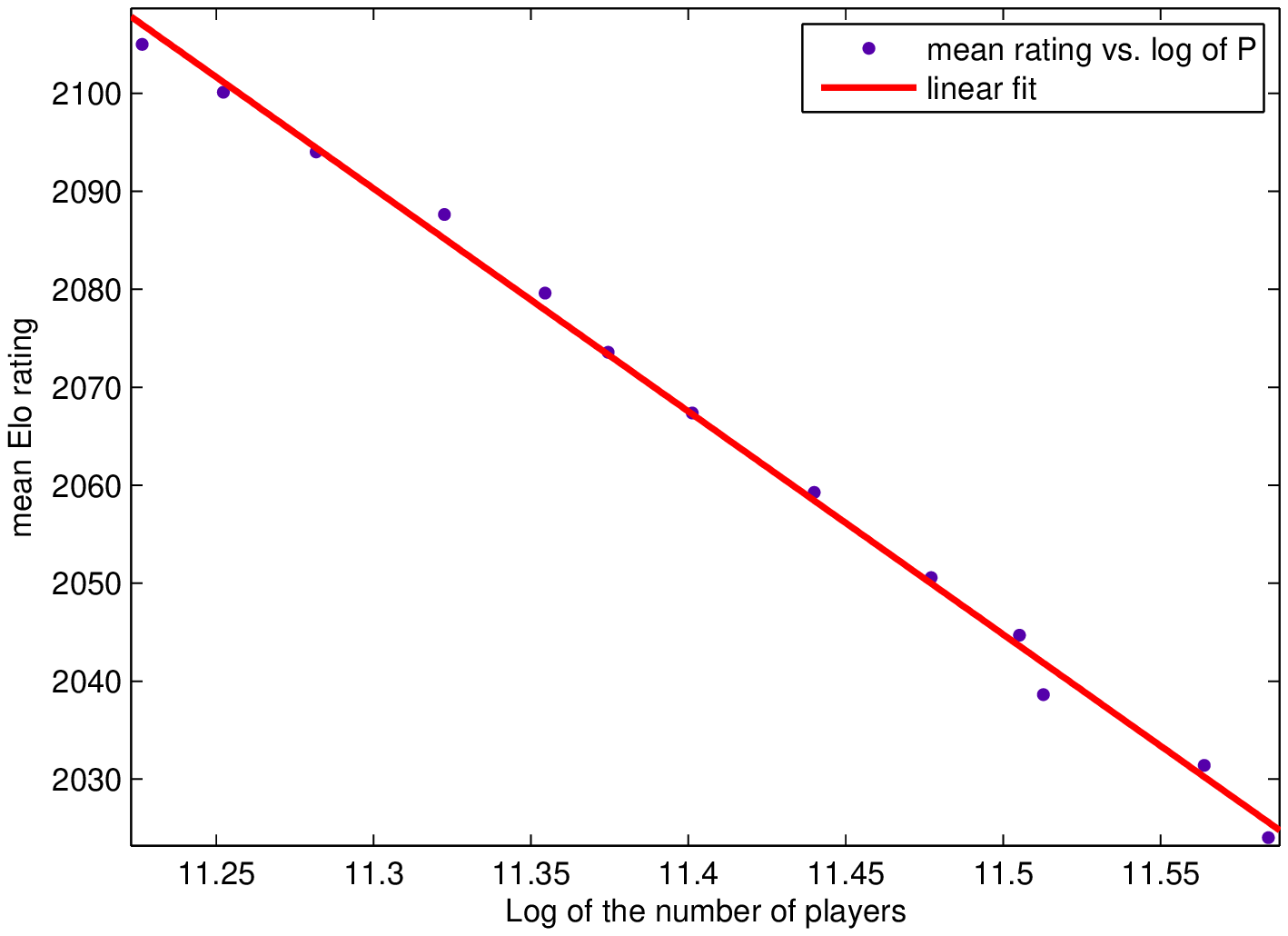} } \\
\end{minipage}
\vspace*{8pt}\caption{\label{fig:mean-skew} Skewness of rating (left) and mean rating linear fit (right), for 13 periods from October 2006 to September 2009}
\end{figure}
\smallskip

The skewness of the actual FIDE rating data is shown in the left-hand plot in Figure~\ref{fig:mean-skew}.
As can be seen, it shows that there is a small negative skew, which has generally slowly increased over the period.
(The increase in skewness in September 2009 is mostly due to FIDE temporarily lowering the minimum rating for new players from 1400 to 1200, and then reverting to the original policy in the following period.) The negative skew can be attributed to the slow decrease in the mean rating with the growing number of players, since it is more likely that a new player joining the pool will enter with a rating lower than the average. This can be formalised as follows.

\smallskip

Let $P_1$ be the number of players in the pool at the end of the first period and let $\mu_1$ be the mean rating of those players. We define $P_2$ and $\mu_2$ similarly for the second period. Then the total of the ratings of all players in the pool is $P_1 \mu_1$ for the first period and $P_2 \mu_2$ for the second period.
Assuming the average rating of new players joining during the second period is $\mu_1 - \epsilon$, we have
\begin{displaymath}
P_2 \mu_2 = P_1 \mu_1 + (P_2 - P_1) (\mu_1 - \epsilon),
\end{displaymath}
yielding
\begin{equation}\label{eq:extra}
\mu_2 - \mu_1 =  - \frac{\epsilon (P_2 - P_1)}{P_2}.
\end{equation}

We can approximate (\ref{eq:extra}) by the differential equation
\begin{displaymath}
\frac{d \mu}{d P} =  - \frac{\epsilon}{P},
\end{displaymath}
which has the solution
\begin{equation}\label{eq:log-nump}
\mu = \mu_0 - \epsilon \ln P,
\end{equation}
where $\mu_0$ is a constant.

\smallskip

The right-hand graph in Figure~\ref{fig:mean-skew} shows the mean Elo rating $\widehat{\mu}$ plotted against the logarithm of the number of players $P$. The linear fit shown is in good agreement with (\ref{eq:log-nump}), with $\epsilon = 227.6$, $\mu_0 = 4663$,  and $R^2 = 0.9964$. Thus the average rating is decreasing slowly as a linear function of the logarithm of the number of players in the pool. In addition, knowing $\epsilon$ would allow us to predict the rate of decrease, and also to estimate the skewness shown in the left-hand graph in Figure~\ref{fig:mean-skew}.

\smallskip

As we have seen above, the Gaussian distribution is a good first approximation. We pursue this further in Section \ref{sec:derive} after we formalise the evolutionary  model for players' ratings in Section~\ref{sec:urn}. We return in Section~\ref{sec:skew} to a more general model that takes skewness into account.

\section{An Evolutionary Urn Transfer Model}
\label{sec:urn}

In our evolutionary stochastic model for rating game players, two main types of event may take place.
The first event type occurs  when a new player enters the system.  We make two assumptions related to such an event:
\renewcommand{\labelenumi}{(\roman{enumi})}
\begin{enumerate}
\item that new players enter the system at a fixed rate, and

\item that once players enter the system they do not leave it.
\end{enumerate}

(We note that the model can be extended to allow players to leave the pool as long as the rate at which players enter the pool is greater than the rate at which they leave.)

The second event type occurs when a game is played between two players. In this case, we assume
\renewcommand{\labelenumi}{(\roman{enumi})}
\begin{enumerate} \setcounter{enumi}{2}
\item that the outcome of the game is either a win or loss for the first player, and

\item that every game occurs between two players of fairly similar strength; in particular, we assume that the absolute value of the difference in strength between the players in any game is at most $W$.
\end{enumerate}

Assumption (iii) is often made, cf. \cite{GLIC99a}, to avoid including extra parameters in the model, as it is reasonable to assume that a draw is equivalent to half a win and half a loss (which is consistent with the score of a draw being $0.5$, as in Section~\ref{sec:elo}); see \cite{JOE91,HENE92,GLIC99b} for alternative ways of dealing with draws.
The basis for Assumption (iv) is that players will normally play games against players of comparable strength; for example, many tournaments are divided into separate grading sections for that reason.
We note that the win probabilities given by (\ref{eq:log}) satisfy
\begin{equation}\label{eq:one}
p_{\alpha \beta} + p_{\beta \alpha} = 1,
\end{equation}
which is consistent with Assumption (iii).

\smallskip

In our model, we approximate the ratings using a discrete numerical scale of values at intervals of $I$. We use {\em urns}  to store the pool of players, with each urn containing players of approximately similar strength. Let $M$ denote the average rating of all the players. Then $urn_k$, the $k$th urn, where $-\infty < k < \infty$, contains those players whose rating is in the range $[M+(k - 0.5)I , M+(k+ 0.5)I)$, i.e. the players are grouped into bins of {\em width} $I$. Thus a player with rating $R$ will be in urn number $\lfloor 0.5 + (R-M)/I \rfloor$.

\smallskip

Players enter the system at a rate $r$, where $0 < r < 1$. After playing a game, a player may stay in the same urn or be transferred to one of the two neighbouring urns, depending on the result of the game. We now describe the urn model in detail.

\smallskip

We assume a countable number of urns, with $urn_0$ being the central urn; to its left are the urns with negative subscripts and to its right are the urns with positive subscripts. We let $F_i(t)$ denote the number of players in $urn_i$ at stage $t$ of the stochastic process. Initially $t=0$, $F_0(0) = \Delta$, with $\Delta > 0$, i.e. $urn_0$ initially has $\Delta$ players in it, and all other urns are empty, i.e. $F_i(0) = 0$ for $i \neq 0$.

\smallskip

When a player enters the system, an existing player $A$ is selected uniformly at random from the urns and the new player is put into the same urn as player $A$, i.e. we assign the new player the same approximate rating as the selected existing player $A$. In other words, new players enter the system according to the distribution of players currently in the system.

\smallskip

The stochastic process modelling the changes in rating can be viewed as a random walk \cite{RUDN04}, where the probabilities of players increasing, decreasing or maintaining their ratings depend on their current ratings, as explained below.

\smallskip

At time $t$, $t > 0$, a player $A$ is chosen uniformly at random from the urns, say from $urn_i$, i.e. $urn_i$ is selected with probability
\begin{equation}\label{eq:pchoice}
\frac{F_i(t)}{\sum_{j = -\infty}^\infty F_j(t)} \approx \frac{F_i(t)}{r t + \Delta},
\end{equation}
where $\approx$ means {\em is approximately equal to for large t}. (This approximation holds since the expectation of the number of players $P$ at time $t$ is $ rt + \Delta$.)

\medskip

As above, we assume $0 < r < 1$. Then one of two things may occur:
\renewcommand{\labelenumi}{(\Roman{enumi})}
\begin{enumerate}
\item with probability $r$, a new player is inserted into $urn_i$, i.e. into the same urn as the chosen player $A$;

\item with probability $1-r$, an opponent $B$ for the player $A$ is chosen from urns
\begin{displaymath}
urn_{i-w}, urn_{i-w+1}, \ldots, urn_{i-1}, urn_i, urn_{i+1}, \ldots, urn_{i+w-1}, urn_{i+w},
\end{displaymath}
where $w = \lfloor W/I \rfloor$.

The probability that player $B$ is chosen from $urn_{i+s}$ is $\pi_s$, $-w \le s \le w$, where for symmetry we assume $\pi_{-s} = \pi_s$. Depending on the result of the game, player $A$ either moves to $urn_{i-1}$ or $urn_{i+1}$, or remains in $urn_i$. The probabilities of these events are chosen so that the expected change in $A$'s rating is identical to that prescribed by the Elo system.
\end{enumerate}
\smallskip

As we are working in terms of urn numbers rather than Elo ratings, we let $c = C I$, so $c$ is the scaling factor in terms of urn numbers. Thus, since $L_C(s I) = L_c(s)$ and $L_C(- s I) = L_c(- s)$, the probability that player $A$ wins is $L_c(-s)$, by (\ref{eq:log}). Therefore, from (\ref{eq:adjust}) and (\ref{eq:sum}), when $A$ wins $A$'s new rating is given by
\begin{equation}\label{eq:new1}
new \ R_A = old \ R_A + K (1 - L_c(-s)) = old \ R_A + K L_c(s).
\end{equation}

In order to find the new urn number $new \ i_A$ for $A$, corresponding to the rating $new \ R_A$, we first normalise (\ref{eq:new1}) by subtracting $M$ and dividing  by $I$, giving
\begin{displaymath}
new \ i_A = old \ i_A + \left( \frac{K}{I} \right) L_c(s).
\end{displaymath}

We restrict player $A$ to moving up or down by at most one urn.
Moreover, we discretise the change stochastically so that the new urn number will be integral but the expected change unaffected.  Hence,
\begin{equation}\label{eq:new2}
new \ i_A = old \ i_A + \left\{
\begin{array}{r@{\quad \quad}l}
1 & {\rm with \ probability} \left( \frac{K}{I} \right) L_c(s) \\
0 & {\rm otherwise.}
\end{array}
\right.
\end{equation}
We note that $I$ has to be chosen so that the probability in (\ref{eq:new2}) does not exceed 1 for all $s$, $-w \le s \le w$.
We therefore require $K \le I (1 + e^{- c w})$. For simplicity, we will choose $I = K = 20$.

\smallskip

The probability that player $A$ moves to $urn_{i+1}$ is
\begin{displaymath}
L_c(-s) \left( \frac{K}{I} \right) L_c(s),
\end{displaymath}
i.e. the product of the probability that $A$ wins against $B$ and the corresponding discretisation probability.

\smallskip

Similarly, when $A$ loses we have
\begin{displaymath}
new \ i_A = old \ i_A - \left( \frac{K}{I} \right) L_c(-s).
\end{displaymath}
\smallskip

Again restricting $A$ to moving up or down by at most one urn, on stochastically discretising, we obtain
\begin{equation}\label{eq:new3}
new \ i_A = old \ i_A - \left\{
\begin{array}{r@{\quad \quad}l}
1 & {\rm with \ probability} \left( \frac{K}{I} \right) L_c(-s) \\
0 & {\rm otherwise.}
\end{array}
\right.
\end{equation}
\smallskip

Therefore the probability that $A$ moves to $urn_{i-1}$ is
\begin{displaymath}
L_c(s) \left( \frac{K}{I} \right) L_c(-s),
\end{displaymath}
i.e. the product of the probability that $A$ loses against $B$ and the corresponding discretisation probability.

\smallskip

Let
\begin{equation}\label{eq:psis}
\psi_s = \left( \frac{K}{I} \right) L_c(s) L_c(-s) =  \left( \frac{K}{I} \right) L_c(s) \left( 1-L_c(s) \right).
\end{equation}
\smallskip

Then, in summary, if the selected player $A$ is from $urn_i$ and the chosen opponent $B$ is from $urn_{i+s}$, $-w \le s \le w$, \renewcommand{\labelenumi}{(\roman{enumi})}
\begin{enumerate}
\item with probability $\psi_s$ player $A$ moves to $urn_{i+1}$,

\item with probability $\psi_s$ player $A$ moves to $urn_{i-1}$, and

\item with probability $1 -2\psi_s$ player $A$ stays in $urn_{i}$.
\end{enumerate}
\medskip

We note that $\psi_s$ is proportional to the derivative of the logistic function, viz.
\begin{displaymath}
\psi_s = \left( \frac{K}{c I} \right)  {L_c}^\prime (s).
\end{displaymath}
\smallskip

This symmetric bell-shaped curve is proportional to the probability density function of the logistic distribution, with  standard deviation $\pi / c \sqrt{3}$ \cite{EVAN00}.

\medskip

It is easy to show that, conditional on $A$ being chosen from $urn_i$ and $B$ from $urn_{i+s}$, the variance of the change in rating is $2 \psi_s$, whereas with the Elo system it is only $(K/I) \psi_s$; the additional variance is due to the stochastic discretisation. It therefore follows that the unconditional variance in our model will also be increased by a factor of $2I/K$  compared to that for the Elo system.

\smallskip

It is clear that, according to the Elo model, player $B$'s rating should be updated in a similar manner to player $A$'s.
However, we simplify the analysis by considering each game as essentially equivalent to two ``half games'', since the players are chosen randomly. It is therefore sufficient to analyse only the change to $A$'s rating.

\smallskip

(We note that, unlike the proposal in \cite{GLIC99b}, our evolutionary model does {\em not} take into account, for example, the fact that junior players tend to be under-rated and to improve more rapidly than older players.)

\section{Derivation of the Distribution of Players' Ratings}
\label{sec:derive}

Considering all possible choices for player $B$, it follows from the above discussion that the probability $\theta$ that $A$ will move to $urn_{i+1}$ is given by
\begin{equation}\label{eq:move1}
\theta = \sum_{s=-w}^{w} \pi_s \psi_s
\end{equation}
and, by symmetry, that this is also the probability that $A$ will move to $urn_{i-1}$.

\medskip

At time $t$, $t > 0$, a game is played with probability $1-r$, and there are then the following three possible ways that the contents of $urn_i$ may change.
\renewcommand{\labelenumi}{(\alph{enumi})}
\begin{enumerate}
\item The player $A$ chosen uniformly at random is selected from $urn_i$, and then plays an opponent $B$ from say $urn_{i+s}$. By (\ref{eq:move1}), the probability that $A$ beats $B$ and moves to $urn_{i+1}$ is $\theta$, that $A$ loses to $B$ and moves to $urn_{i-1}$ is $\theta$, and that $A$ stays in $urn_i$ is $1 - 2 \theta$. Thus the net expected loss from $urn_i$ is $2 \theta$.

\item The player $A$ chosen uniformly at random is selected from $urn_{i-1}$, and then plays an opponent $B$ from say $urn_{i-1+s}$. By (\ref{eq:move1}), the probability that $A$ beats $B$ and moves to $urn_i$ is $\theta$; so the net expected gain to $urn_i$ is $\theta$. (In all other cases the contents of $urn_i$ do not change.)

\item The player $A$ chosen uniformly at random is selected from $urn_{i+1}$, and then plays an opponent $B$ from say $urn_{i+1+s}$. By (\ref{eq:move1}), the probability that $A$ loses to $B$ and moves to $urn_i$ is $\theta$; so the net expected gain to $urn_i$ is $\theta$. (In all other cases the contents of $urn_i$ do not change.)
\end{enumerate}
If $A$ is selected from any of the other urns, the contents of $urn_i$ do not change.

\smallskip

We now obtain the difference equation for the urn transfer model, by considering the expected change to $urn_i$, as discussed above. For integer $i$ and $t \ge 0$,
\begin{equation}\label{eq:walk}
E(F_i(t+1)) = F_i(t) + \frac{1-r}{P} \left( \theta F_{i-1}(t) + \theta F_{i+1}(t) - 2 \theta F_{i}(t) \right) + \frac{r}{P} F_i(t).
\end{equation}
\smallskip

To derive (\ref{eq:walk}), we follow a mean-field theory approach, such as that in \cite{OPPE01,LEVE01c},
replacing $P$ by its expectation $rt + \Delta$, as in (\ref{eq:pchoice}). The expected value of $F_i(t+1)$ is equal to the previous number of players in $urn_i$ plus the two probabilities of inserting a player into $urn_i$, from either $urn_{i-1}$ or $urn_{i+1}$, minus the probability of moving a player from $urn_i$ to either of the neighbouring urns, i.e. $urn_{i-1}$ and $urn_{i+1}$, plus the probability of inserting a new player into $urn_i$.

\smallskip

We now take expectations in (\ref{eq:walk}), and we write $F_i(t)$ for $E(F_i(t))$. By the linearity of $E(\cdot)$, we obtain
\begin{equation}\label{eq:expected-walk}
F_i(t+1) - F_i(t) = \frac{\theta(1-r)}{r t + \Delta} \left( F_{i-1}(t) + F_{i+1}(t) - 2 F_{i}(t) \right) + \frac{r}{r t + \Delta} \ F_i(t).
\end{equation}

We note that (\ref{eq:expected-walk}) defines a symmetric random walk by the selected player $A$ at time $t$, where the probability of moving right or left is proportional to $\theta$, but the probability that $A$ is selected decreases over time. Thus the distribution of the players in the urns flattens asymptotically over time and the standard deviation increases, as in a diffusion process \cite{DILL03}.

\smallskip

We will see that in our case the variance increases logarithmically with time and thus the distribution will flatten very slowly.

\medskip

We now approximate our discrete model by a continuous model using a continuous function $F(i,t)$ to approximate $F_i(t)$. In particular, we may approximate
\begin{displaymath}
F_i(t+1) - F_i(t) \quad {\rm by} \quad \frac{\partial F(i,t)}{\partial t}
\end{displaymath}
and
\begin{displaymath}
F_{i-1}(t) + F_{i+1}(t) - 2 F_i(t) \quad {\rm by} \quad \frac{\partial^2 F(i,t)}{\partial i^2}.
\end{displaymath}
From (\ref{eq:expected-walk}), we thus derive the partial differential equation
\begin{equation}\label{eq:partial-f}
\left( t + \frac{\Delta}{r} \right) \frac{\partial F(i,t)}{\partial t} =  \lambda \frac{\partial^2 F(i,t)}{\partial i^2} + F(i,t),
\end{equation}
where
\begin{equation}\label{eq:lambda}
\lambda = \frac{\theta (1-r)}{r}
\end{equation}
is a constant. 

\smallskip

If we now let
\begin{displaymath}
F(i,t) = (rt + \Delta) G(i,t),
\end{displaymath}
we can transform (\ref{eq:partial-f}) into
\begin{equation}\label{eq:partial-f-tau}
\left(t + \frac{\Delta}{r}  \right) \frac{\partial G(i,t)}{\partial t} =  \lambda \frac{\partial^2 G(i,t)}{\partial i^2}.
\end{equation}

\smallskip

We now transform (\ref{eq:partial-f-tau}) into the following simple form of the {\em standard diffusion equation}
(also known as the {\em heat equation}) \cite{DILL03,RUDN04}, by making the substitution $t = \Delta (e^z -1)/r$ and writing $H(i,z)$ for $G(i,t)$:
\begin{equation}\label{eq:heat1}
\frac{\partial H(i,z)}{\partial z} = \lambda \frac{\partial^2 H(i,z)}{\partial i^2}.
\end{equation}

The initial conditions of the discrete model are $F_0(0) = \Delta$, where $\Delta > 0$, and $F_i(0) = 0$ for $i \not= 0$. Since
\begin{displaymath}
\sum_{i = - \infty}^{\infty} F_i(0) = \Delta,
\end{displaymath}
the boundary conditions for the continuous model become $F(i,0) = \Delta \delta(i)$, where $\delta(\cdot)$ is the Dirac delta function. This yields the boundary conditions

\begin{equation}\label{eq:boundary}
H(i,0) = G(i,0) = \frac{F(i,0)}{\Delta} = \delta(i).
\end{equation}

Equation (\ref{eq:heat1}) with boundary conditions (\ref{eq:boundary}) has the following standard solution:
\begin{equation}\label{eq:heat2}
H(i,z) = \frac{1}{\sqrt{4 \pi \lambda z}} \exp \left( \frac{- i^2}{4 \lambda z} \right),
\end{equation}
and we see from (\ref{eq:gauss}) that this is the density function of the Gaussian distribution with mean 0 and variance $2 \lambda z$.

From (\ref{eq:heat2}) it follows that
\begin{equation}\label{eq:heat3}
F(i,t) = \frac{rt + \Delta}{[4 \pi \lambda \ln (1+\frac{r t}{\Delta})]^{1/2}} \exp \left( \frac{- i^2}{4   \lambda \ln (1 + \frac{r t}{\Delta})} \right).
\end{equation}

\section{Modelling the Distribution of Chess Players' Ratings}
\label{sec:chess}

In order to run simulations of the model that we described and analysed in Sections \ref{sec:urn} and \ref{sec:derive}, respectively, we first need to specify or derive values for the various parameters involved.

We are assuming that $C = \ln 10 / 400$ and $I = K = 20$, as stated previously in Sections \ref{sec:elo} and \ref{sec:urn}; thus $c = \ln 10 / 20$. We consider the cases $w = 1,2$ and $3$, and for simplicity we assume that the urn from which the opponent $B$ is selected is chosen uniformly, i.e. $\pi_s = 1/(2w +1)$. We can then compute $\psi_s$ from (\ref{eq:psis}) and $\theta$ from (\ref{eq:move1}).

\smallskip

Finally, we need estimates for $r$, $\Delta$ and $t$.
We assume, as indicated in Section~\ref{sec:fide}, that the ratings are normally distributed; we relax this assumption in Section~\ref{sec:skew} to cater for some degree of skewness in the distribution.
In order to validate our model, we obtain estimates for these parameters using the published official rating data from January 2007 to January 2010, as described in Section~\ref{sec:fide}. Our methodology is to extract values for these parameters from this data,
using the analysis in Section~\ref{sec:derive},
and then run simulations of our model in order to see how closely the resulting distribution matches that obtained from the actual data.

\smallskip

To estimate $r$ from the actual rating data, we proceed in the following way.
Let $P$ be the number of rated players recorded at January of a given year. Let $G$ be the number of games played
and $N$ be the number of new players joining the pool of rated players during the previous year
(computed as the difference between $P$ and its value for the previous January).
According to the data, the rate $r$ at which players entered the system during the previous year is given by
\begin{displaymath}
r = \frac{N}{N + G}.
\end{displaymath}

The values for these parameters from January 2007 to January 2010, calculated using the official FIDE data, are presented in Table~\ref{table:r}. In the simulations we took the rate $r$ to be $0.009553$, the average rate over the complete four-year period, as shown in the summary row. It can be seen from the table that, in reality, $r$ fluctuates somewhat, but as an approximation we assume that $r$ is constant. We can then compute $\lambda$ from (\ref{eq:lambda}).

\begin{table}[ht]
\begin{center}
\begin{tabular}{|c|c|c|c|c|}\hline
Year    & $P$    & $G$     & $N$   & $r$      \\ \hline
2006    & 67349  & $-$     & $-$   & $-$      \\
2007    & 77056  & 881089  & 9707  & 0.010897 \\
2008    & 87075  & 1009067 & 10019 & 0.009831 \\
2009    & 99223  & 1181206 & 12148 & 0.010180  \\
2010    & 109373 & 1285607 & 10150 & 0.007833 \\ \hline
Summary & $-$ & 4356969 & 42024 & 0.009553    \\ \hline
\end{tabular}
\end{center}
\caption{\label{table:r} The data used to compute $r$}
\end{table}
\smallskip

Lastly we need to obtain values for $\Delta$ and $t$. From (\ref{eq:gauss}) and (\ref{eq:heat3}), it follows that at time $t$ the expected number of players $P$ is given by
\begin{equation}\label{eq:t}
P = r t  + \Delta,
\end{equation}
and that $\sigma^2$, the variance of the rating distribution, is $2 \lambda \ln (1 + rt/\Delta)$. We thus obtain
\begin{equation}\label{eq:delta}
\Delta = P \exp \left(\frac{- \sigma^2 }{2 \lambda} \right).
\end{equation}
\smallskip

To get a single value for $\Delta$, we simply take the average over the years 2007 to 2010, where we compute a year-specific value for $\Delta$ from (\ref{eq:delta}) using the values of $P$ and $\sigma$ from Table~\ref{table:gauss}.
Finally, we estimate $t$ using (\ref{eq:t}).

\smallskip

For $w =1,2$ and $3$, the estimated values for $\Delta$ and $t$ are presented in Table~\ref{table:t}, where the values for $t$ are rounded to the nearest 10. We also obtained alternative estimates by replacing $P$ by $Q$ in (\ref{eq:t}) and (\ref{eq:delta}); the two alternatives are indicated by the first column of Table~\ref{table:t}. The alternatives will be denoted by
$\Delta_P,t_P$ and $\Delta_Q,t_Q$, respectively.

\begin{table}[ht]
\begin{center}
\begin{tabular}{|c|c|c|c|c|c|c|}\hline
Using  & $w$ & $\Delta$ & $t$ until 2007 & $t$ until 2008 & $t$ until 2009 & $t$ until 2010 \\ \hline
$P$    & 1   & 20701    & 5899130        & 6947900        & 8219530        & 9282010 \\
$Q$    & 1   & 20242    & 5749450        & 6762420        & 8042210        & 9173160 \\
$P$    & 2   & 20569    & 5912960        & 6961730        & 8233360        & 9295840 \\
$Q$    & 2   & 20113    & 5762980        & 6775950        & 8055750        & 9186690 \\
$P$    & 3   & 20373    & 5933480        & 6982250        & 8253880        & 9316360 \\
$Q$    & 3   & 19921    & 5783070        & 6796030        & 8075830        & 9206770  \\ \hline
\end{tabular}
\end{center}
\caption{\label{table:t} Derived $\Delta$ and $t$ for 2007 to 2010, for $w=1,2$ and $3$}
\end{table}
\smallskip

As mentioned above, we fixed $r$ at $0.009553$, the value obtained in Table~\ref{table:r}.
For each set of values for the parameters $w$, $\Delta$ and $t$ in Table~\ref{table:t}, we ran 10 simulations of the stochastic process described in Section~\ref{sec:urn}, implemented in Matlab. In each case we then fitted a Gaussian to the distribution of the number of players in the urns, again using Matlab. Each row in Table~\ref{table:simul} was computed from the average of the 10 simulations in exactly the same way that the values in Table~\ref{table:gauss} were computed from the actual rating data. That is, $P$, $\widehat{\mu}$ and $\widehat{\sigma}$ are the values calculated from the results of the simulations, and $Q$, $\mu$ and $\sigma$ are the values obtained by fitting a Gaussian distribution to the simulation results.
(In order to obtain Elo ratings from the urn numbers of the players in the simulation, the urn numbers were calibrated by means of a suitable shift. This was chosen so that the means $\widehat{\mu}$ from Table~\ref{table:gauss} for each of the four years were within the range of $urn_0$.) It can be seen that, in each row of Table~\ref{table:simul}, all the fitted and calculated values are very close to each other. This and the fact that $R^2$ is so close to one gives strong confirmation of our analysis in Section~\ref{sec:derive}.

\smallskip

We now compare the fitted and calculated parameters from Table~\ref{table:simul} with those in Table~\ref{table:gauss}.
Obviously, by construction, $\mu$ and $\widehat{\mu}$ are very close to the corresponding values in Table~\ref{table:gauss}.
In addition, it can be seen that the values for $P$ and $Q$ when using $\Delta_P$ and $t_P$ are very close to the values for $P$ in Table~\ref{table:gauss}, and correspondingly close to the values for $Q$ in Table~\ref{table:gauss} when using $\Delta_Q$ and $t_Q$.
However, the calculated standard deviation $\widehat{\sigma}$ in Table~\ref{table:simul} is consistently lower than its counterpart in Table~\ref{table:gauss}. For 2007 they are very close, for 2008 they are about 10 Elo points apart, for 2009 they are about 17 points apart, while for 2010 they are about 24 points apart.
Although these results are very encouraging, we will see in the next section that we can get much closer to the actual standard deviations by introducing skewness into the model.

\begin{table}[ht]
\begin{center}
\begin{tabular}{|c|c|c|c|c|c|c|c|c|c|}\hline
Using & $w$ & Year & $Q$    & $\mu$   & $\sigma$ & $R^2$  & $P$    & $\widehat{\mu}$ & $\widehat{\sigma}$ \\ \hline
$\Delta_P,t_P$  & 1   & 2007 & 77016  & 2092.231 & 164.476  & 0.9993 & 77064  & 2092.179       & 164.890 \\
$\Delta_Q,t_Q$   & 1   & 2007 & 74895  & 2103.906 & 164.826  & 0.9993 & 74899  & 2104.053       & 164.923 \\
$\Delta_P,t_P$   & 2   & 2007 & 77048  & 2096.596 & 164.865  & 0.9994 & 77045  & 2096.646       & 164.933 \\
$\Delta_Q,t_Q$   & 2   & 2007 & 75206  & 2090.258 & 164.489  & 0.9993 & 75233  & 2090.195       & 164.752 \\
$\Delta_P,t_P$   & 3   & 2007 & 77201  & 2091.796 & 164.679  & 0.9994 & 77228  & 2092.199       & 164.981 \\
$\Delta_Q,t_Q$   & 3   & 2007 & 75066  & 2095.629 & 164.533  & 0.9993 & 75111  & 2096.049       & 164.893 \\ \hline
$\Delta_P,t_P$   & 1   & 2008 & 87049  & 2075.647 & 171.938  & 0.9994 & 87081  & 2075.632       & 172.178 \\
$\Delta_Q,t_Q$   & 1   & 2008 & 84901  & 2077.914 & 172.079  & 0.9994 & 84962  & 2078.193       & 172.515 \\
$\Delta_P,t_P$   & 2   & 2008 & 87080  & 2084.221 & 172.265  & 0.9995 & 87111  & 2083.853       & 172.573 \\
$\Delta_Q,t_Q$   & 2   & 2008 & 84829  & 2069.993 & 172.106  & 0.9994 & 84843  & 2070.036       & 172.171 \\
$\Delta_P,t_P$   & 3   & 2008 & 87073  & 2075.797 & 172.173  & 0.9994 & 87144  & 2075.890       & 172.676 \\
$\Delta_Q,t_Q$   & 3   & 2008 & 84897  & 2070.297 & 171.903  & 0.9994 & 84926  & 2070.029       & 172.086 \\ \hline
$\Delta_P,t_P$   & 1   & 2009 & 99267  & 2033.206 & 179.812  & 0.9995 & 99324  & 2033.229       & 180.227 \\
$\Delta_Q,t_Q$   & 1   & 2009 & 96970  & 2036.315 & 179.620  & 0.9995 & 97030  & 2036.540       & 180.000 \\
$\Delta_P,t_P$   & 2   & 2009 & 99121  & 2037.768 & 179.613  & 0.9995 & 99141  & 2037.980       & 179.862 \\
$\Delta_Q,t_Q$   & 2   & 2009 & 96985  & 2041.688 & 179.421  & 0.9995 & 97043  & 2041.843       & 179.820 \\
$\Delta_P,t_P$   & 3   & 2009 & 99149  & 2032.080 & 179.770  & 0.9995 & 99166  & 2031.835       & 179.943 \\
$\Delta_Q,t_Q$   & 3   & 2009 & 97138  & 2032.044 & 179.542  & 0.9994 & 97172  & 2031.974       & 179.804 \\ \hline
$\Delta_P,t_P$   & 1   & 2010 & 109458 & 2019.797 & 185.313  & 0.9995 & 109485 & 2019.818       & 185.528 \\
$\Delta_Q,t_Q$   & 1   & 2010 & 107868 & 2026.636 & 186.163  & 0.9995 & 107893 & 2026.662       & 186.304 \\
$\Delta_P,t_P$   & 2   & 2010 & 109457 & 2014.156 & 185.612  & 0.9995 & 109469 & 2013.943       & 185.691 \\
$\Delta_Q,t_Q$   & 2   & 2010 & 107847 & 2021.331 & 185.932  & 0.9995 & 107860 & 2021.283       & 186.047 \\
$\Delta_P,t_P$   & 3   & 2010 & 109373 & 2013.765 & 185.203  & 0.9994 & 109386 & 2013.833       & 185.344 \\
$\Delta_Q,t_Q$   & 3   & 2010 & 107840 & 2020.426 & 185.611  & 0.9995 & 107865 & 2020.262       & 185.806 \\ \hline
\end{tabular}
\end{center}
\caption{\label{table:simul} Actual and fitted parameters for simulation results}
\end{table}

\section{Taking Skewness into Account}
\label{sec:skew}

As discussed in Section~\ref{sec:fide}, the actual rating data exhibits a small negative skew. We now consider modifying the urn model
presented in Section~\ref{sec:urn} to take this into account.
Since it is likely that a new player will enter with a rating lower than the average, we can model this skewness in a simple way by making a small change to the way in which new players are added.
Instead of inserting the new player into the same urn as the chosen player $A$, say $urn_i$, we put the new player into $urn_{i-\kappa}$, where $\kappa$ determines the amount of negative skew we wish to introduce.

\smallskip

To validate the modified stochastic process, we ran a batch of simulations in Matlab, starting the process with the actual rating data as of October 2006 and ending in January 2010. For the October 2006 starting data $\widehat{\mu} = 2105.007$, $\widehat{\sigma} = 163.552$ and $s = -0.1354$ (as shown in Figure~\ref{fig:mean-skew}).
From October to December 2006 the number of games played was $G = 259,662$, and the number of new players was $N = 1960$.
Using these values together with the data in Table~\ref{table:r}, we therefore took the number of simulation steps to be $N+G = 3,769,819$ and, as before, the rate $r$ at which players enter the system to be $0.009553$.
Tables \ref{table:k1}, \ref{table:k2} and \ref{table:k3} show the average skewness $s$, mean rating $\widehat{\mu}$ and standard deviation $\widehat{\sigma}$ over 10 simulations, for $w = 1,2$ and $3$, respectively, with $\kappa$ varying from $0$ to $12$. As a reference point, for the actual rating data as of January 2010, $\widehat{\mu} = 2015.650$ and $\widehat{\sigma} = 209.622$, as in Table~\ref{table:gauss}, and we computed $s = -0.2284$.

It can be seen that the results are rather similar in all three tables.
As $\kappa$ is increased, the skewness $s$ becomes more negative, the mean $\widehat{\mu}$ decreases and the standard deviation
$\widehat{\sigma}$ increases, as expected.
The closest fit to the actual skewness $s = -0.2284$ and the standard deviation $\widehat{\sigma} = 209.622$ is when $\kappa$ is $8$.
However, the closest fit to the mean Elo rating $\widehat{\mu} = 2015.650$ is when $\kappa$ is $11$ or $12$.
The suggested values for $\kappa$ therefore correspond to a new player being rated $160-230$ Elo points below the average rating.
This latter value is in broad agreement with the value $\epsilon = 227.6$ obtained in Section~\ref{sec:fide} from Figure~\ref{fig:mean-skew}. Although this value was obtained using the entire three year period, the values for the individual years calculated from (\ref{eq:extra})
are similar, being roughly in the range $200-300$.
These results confirm that the modified process is a reasonable model for obtaining rating data with the observed parameters, despite the discrepancy between the values for $\kappa$.
This discrepancy is not surprising, since the modified model, as a first approximation, is clearly an oversimplification.
We note that the value of $w$ seems to have very little effect on the results, although it is possible that some pattern might be noticeable if a significantly larger value for $w$ was used.


\begin{table}[ht]
\begin{center}
\begin{tabular}{|c|c|c|c|}\hline
$\kappa $ & $s$ & $\widehat{\mu}$ & $\widehat{\sigma}$ \\ \hline
$-$& {\em -0.2284}  & {\em 2015.650}  & {\em 209.622} \\ \hline
0  & -0.0920 & 2105.641	& 186.980 \\
1  & -0.0884 & 2097.907	& 187.030 \\
2  & -0.0933 & 2089.894	& 188.557 \\
3  & -0.0995 & 2082.068	& 190.721 \\
4  & -0.1064 & 2074.321	& 193.440 \\
5  & -0.1276 & 2066.435	& 197.046 \\
6  & -0.1541 & 2058.940 & 201.387 \\
7  & -0.1881 & 2050.446	& 206.373 \\
8  & {\bf -0.2309} & 2043.175	& {\bf 211.843} \\
9  & -0.2745 & 2035.115	& 218.153 \\
10 & -0.3206 & 2027.404	& 224.623 \\
11 & -0.3742 & {\bf 2019.250}	& 232.101 \\
12 & -0.4227 & {\bf 2012.005}	& 239.422 \\ \hline

\end{tabular}
\end{center}
\caption{\label{table:k1} Simulation results allowing skewness, for $w = 1$}
\end{table}
\begin{table}[ht]
\begin{center}
\begin{tabular}{|c|c|c|c|}\hline
$\kappa $ & $s$ & $\widehat{\mu}$ & $\widehat{\sigma}$ \\ \hline
$-$& {\em -0.2284}  & {\em 2015.650}  & {\em 209.622} \\ \hline
0  & -0.0907 & 2105.510	& 186.788 \\
1  & -0.0917 & 2097.558	& 187.265 \\
2  & -0.0963 & 2089.933	& 188.447 \\
3  & -0.1008 & 2082.062	& 190.765 \\
4  & -0.1114 & 2074.058	& 193.490 \\
5  & -0.1305 & 2066.190	& 197.117 \\
6  & -0.1532 & 2058.516	& 201.448 \\
7  & -0.1845 & 2050.812	& 206.251 \\
8  & {\bf -0.2297} & 2043.222	& {\bf 212.007} \\
9  & -0.2744 & 2035.224	& 217.962 \\
10 & -0.3226 & 2027.335	& 224.759 \\
11 & -0.3722 & {\bf 2019.737}	& 231.852 \\
12 & -0.4346 & {\bf 2011.701}	& 239.575 \\ \hline
\end{tabular}
\end{center}
\caption{\label{table:k2} Simulation results allowing skewness, for $w = 2$}
\end{table}
\begin{table}[ht]
\begin{center}
\begin{tabular}{|c|c|c|c|}\hline
$\kappa $ & $s$ & $\widehat{\mu}$ & $\widehat{\sigma}$ \\ \hline
$-$& {\em -0.2284}  & {\em 2015.650}  & {\em 209.622} \\ \hline
0  & -0.0931 & 2105.713	& 186.641 \\
1  & -0.0905 & 2097.470	& 187.274 \\
2  & -0.0929 & 2090.024	& 188.341 \\
3  & -0.1014 & 2082.383	& 190.398 \\
4  & -0.1105 & 2074.388	& 193.320 \\
5  & -0.1230 & 2066.305	& 196.642 \\
6  & -0.1571 & 2058.612	& 201.401 \\
7  & -0.1860 & 2050.754	& 206.245 \\
8  & {\bf -0.2310} & 2042.906	& {\bf 211.766} \\
9  & -0.2749 & 2035.191	& 217.848 \\
10 & -0.3249 & 2027.141	& 224.830 \\
11 & -0.3716 & {\bf 2019.805}	& 231.613 \\
12 & -0.4290 & {\bf 2011.446}	& 239.475 \\ \hline
\end{tabular}
\end{center}
\caption{\label{table:k3} Simulation results allowing skewness, for $w = 3$}
\end{table}

\section{Concluding Remarks}
\label{sec:conc}


We have constructed a stochastic evolutionary urn model that generates the distribution of players' ratings and have validated this model using published official rating data on chess players. For the symmetric case, our analysis of the model yielded
a Gaussian distribution, which has the interesting feature that the variance increases logarithmically with time. This implies that the distribution of ratings is quite stable, but has the tendency to flatten extremely slowly over time. These results were validated by simulating the model. Although the data is well approximated by a Gaussian, there is a small negative skew present in the data. An improvement can be made to the model to account for this by breaking the symmetry and putting new players into lower-numbered urns, corresponding to new players generally having lower than average ratings. The modified stochastic process was validated by simulation starting with actual rating data. Deriving analytically the distribution for the modified process remains an open problem.

\smallskip


Throughout the paper we have assumed that the $K$-factor is fixed at 20. It would be interesting to allow the $K$-factor to vary with players' ratings and the number of games they have played, as suggested in \cite{GLIC99b}, and to see whether such a modification could shed some light on the $K$-factor controversy mentioned in Section~\ref{sec:elo}.

\newcommand{\etalchar}[1]{$^{#1}$}

\end{document}